\documentclass[twocolumn]{aastex701}

\usepackage{amsmath}
\usepackage{enumerate}

\def\arcsec{{\mbox{$^{\prime\prime}$}}}

\def\Mstellar{{M$_{*}$}}
\def\Msun{{M$_{\odot}$}}

\shorttitle{TiNy Titans: Isolated Compact Dwarf Group}
\shortauthors{TiNy Titans}
\accepted{to ApJ on November 25, 2025}

\begin{document}
\title{TiNy Titans HI: Discovering Satellites via HI Gas in an Isolated, Compact Group of Dwarf Galaxies}
\author[orcid=0000-0002-7532-3328,sname='Stierwalt']{Sabrina Stierwalt}
\affiliation{Department of Physics, Occidental College, 1600 Campus Road, Los Angeles, CA 90041, USA}
\email{sabrina@oxy.edu} 
\author[orcid=0009-0006-6641-0928]{Nicholas Luber}
\affiliation{Department of Astronomy, Columbia University, Mail Code 5247, 538 West 120th Street, New York, NY 10027, USA}
\email{nicholas.m.luber@gmail.com}
\author[]{Zelie Goldberg Little}
 \affiliation{Department of Physics, Occidental College, 1600 Campus Road, Los Angeles, CA 90041, USA}
\email{goldberglitt@oxy.edu}
\author[orcid=0000-0003-3474-1125]{George C. Privon}
\affiliation{National Radio Astronomy Observatory, 520 Edgemont Road, Charlottesville, VA 22903, USA}
\email{gprivon@nrao.edu}
\author[0000-0003-0715-2173]{Gurtina Besla}
\affiliation{Department of Astronomy, University of Arizona, 933 North Cherry Avenue, Tucson, AZ 85721}
\email{gbesla@email.arizona.edu}
\author[0000-0001-8348-2671]{Kelsey E. Johnson}
\affiliation{Department of Astronomy, University of Virginia, 530 McCormick Road, Charlottesville, VA 22904 USA}
\email{kej7a@virginia.edu}
\author[0000-0002-3204-1742]{Nitya Kallivayalil}
\affiliation{Department of Astronomy, University of Virginia, 530 McCormick Road, Charlottesville, VA 22904 USA}
\email{njk3r@virginia.edu}
\author[orcid=0000-0002-1871-4154]{David R. Patton}
\affiliation{Department of Physics \& Astronomy, Trent University, 1600 West Bank Drive, Peterborough, Ontario, Canada K9L OG2}
\email{dpatton@trentu.ca}
\author[0000-0002-1129-1873]{Mary Putman}
\affiliation{Department of Astronomy, Columbia University, Mail Code 5247, 538 West 120th Street, New York, NY 10027, USA}
\email{mep2157@columbia.edu}
\author[0009-0009-7873-3777]{David Simpson Heil}
 \affiliation{Department of Physics, Occidental College, 1600 Campus Road, Los Angeles, CA 90041, USA}
\email{davidsimpsonheil@gmail.com} 


\begin{abstract}
We report on the HI content of an isolated, compact group of 6 dwarf galaxies at a distance of 145 Mpc. The distribution and kinematics of the HI, including multiple gaseous bridges, indicate the group is a gravitationally bound system. The HI maps further reveal two newly discovered dwarf satellites easily identified by their gas but only barely visible in optical images. The four dwarf group members previously identified in SDSS have 9.06 $<$ log(M$_*$/M$_{\odot}$) $<$ 9.43 and 9.42 $<$ log(M$_{HI}$/M$_{\odot}$) $<$ 9.73. The two newly discovered dwarf satellites have log(M$_*$/M$_{\odot}$) $=$ 6.10 with log(M$_{HI}$/M$_{\odot}$) $=$ 8.71 and log(M$_*$/M$_{\odot}$) $=$ 7.07 with log(M$_{HI}$/M$_{\odot}$) $=$ 9.18. New Gemini optical spectra link the HI detections and their optical counterparts. The group's 3D velocity dispersion (188 km/s), mass-to-light ratio (M$_L$/B $\sim$44), dynamical-to-baryonic mass ratio (M$_{dyn}$/M$_{bary}$ $\sim$21), size (69 kpc), and gas fraction (0.56) are all consistent with the compact dwarf groups in the TNG50 simulation. The group has a top-heavy satellite mass function that is inconsistent with predictions for LMC-sized hosts and may instead be two or more groups coming together. A Voronoi tessellation reveals the group resides in a tendril outside the intersection of two filaments. These intermediate density environments within large scale structure provide the conditions needed for groups of star forming, gas-rich dwarf galaxies to form and eventually merge. Our results further show that it is possible to uncover fainter dwarf satellites around dwarf galaxy hosts via HI maps.
\end{abstract}


\section{Introduction}
Under the $\Lambda$-Cold Dark Matter framework, hierarchical structure formation - the merging together of smaller systems to form larger ones - is predicted to happen at even the lowest masses. In simulations, dwarf galaxies themselves host even smaller dark matter subhaloes, and the most massive of those subhaloes are expected to undergo star formation resulting in observable dwarf galaxy pairs and even groups \citep[e.g.,][]{wheeler15,dooley17,wrightRomulus}. 

A growing number of dwarf groups have been observed due to investigations of the satellite mass function around dwarf hosts. Studies have begun to pursue deep optical imaging to look for satellites around dwarf galaxies in the Local Volume outside the Milky Way's immediate influence \citep{carlinMADCASH, delvemcnanna, garling21, davis21, carlinMADCASH24, dolivaMADCASH, medoffMADCASH, LiELVES}. Systematic searches around these very nearby dwarfs (2-10 Mpc) typically probe stellar masses down to $\sim$10$^5$M$_{\odot}$ and identify $\lesssim$1-2 lower mass satellites per dwarf host. The resulting mass functions tend to sit at the low end of predictions from $\Lambda$CDM \citep[e.g.,][]{dooley17, manwadkar, santosSMF}.

Detailed studies of the orbital motions of our nearest neighbors, the Large and Small Magellanic Clouds, suggest they entered into the Milky Way halo as a group and host their own smaller dwarf satellites \citep{dhongialake, BeslaLMC}. Despite this clear nearby example, however,  
theory predicts that, at the current epoch, bound groups with multiple LMC/SMC analogs should be observable but rare \citep[e.g.,][]{guo11,BeslaTNT,manwadkar,santosSMF,katie24}. For example, for galaxies with M$_*\sim$10$^9$M$_{\odot}$, Millennium-II predicts 1-3\% have a companion of similar stellar mass \citep{sales13} and only 4-8\% have a companion with M$_*\sim$10$^8$M$_{\odot}$. In mock catalogs drawn from the low redshift volume of the Illustris simulation, only 3-4\% of dwarfs isolated from massive companions are part of a pair \citep{BeslaTNT}. Setting a mass cutoff for companions of M$_* > $2$\times$10$^8$M$_{\odot}$ and a separation limit of v$_{sep}$ $<$ 150 km/s, \cite{BeslaTNT} further find that groups of more than three dwarfs are found to be ``cosmologically improbable''.

However, hydrodynamic simulations that focus specifically on compact ($<$100 kpc) dwarf groups, find groups with multiple LMC/SMC analogs to exist at the current epoch. They are produced in TNG50-1 simulations that probe stellar masses down to M$_* \sim$ 10$^7$M$_{\odot}$ \citep{dgTNG50}. These groups are typically formed recently (in the past 1.5 Gyr) and eventually merge into one more massive galaxy in a similar time scale (1-3 Gyr).

In a clear example of hierarchical structure formation at low masses, \cite{Stierwalt2017} discovered the first sample of isolated dwarf galaxy groups (3 or more members; $>$1.5 Mpc from a massive neighbor) hosted by LMC/SMC analog pairs. The isolated groups were originally identified in the TiNy Titans (TNT) sample of dwarf pairs 
\citep[M$_* < $10$^{9.5}$M$_{\odot}$;][]{Stierwalt2015} found via the Sloan Digital Sky Survey (SDSS), and their associations were later confirmed via a combination of long-slit spectroscopic observations and narrowband H$\alpha$ imaging \citep{Stierwalt2017}. The TNT groups are distinct from any previously known groups due to their clear isolation from a massive host and their compactness ($<$80 kpc in extent). 

However, group membership on the basis of line-of-sight velocity differences alone is prone to contamination. When comparing to mock catalogs from Illustris, \cite{BeslaTNT} found that 40\% of dwarf multiples identified in observations from SDSS were likely to be projection effects even when line-of-sight velocities were known. In five different semi-analytic models, \citep{taverna22} also found 40\% of the identified groups were actually chance alignments along the line of sight.

Neutral hydrogen is a powerful tracer of galaxy-galaxy interactions and thus can rule out the possibility of chance projection. The HI component of a galaxy typically extends well beyond the optical disk, making it more susceptible to tidal forces \citep[e.g.,][]{hibbard00, interacthikor1,interactHIkor2,  Martin2021}. Specifically, gaseous tidal tails and bridges are signs of ongoing gravitational interactions \citep{toomre1972galactic, bh92, YunM81, varenius, mightee} as are a common, diffuse gas envelope \citep{4485HI, luber22}. Interactions between the LMC and SMC themselves are suspected to have created the gaseous bridge that connects them \citep{putman03, zivick}. Previously unknown companions, specifically optically-faint yet gas-rich dwarfs, can also be discovered via their gas content \citep{Stierwalt2009,shield,mcquinn21, mightee}.

Here we investigate the HI content of the isolated, compact, dwarf group dm1623$+$15  at a distance of 145 Mpc. This group is one of two groups found in the TNT sample of interacting dwarf pairs that has at least four group members \citep{Stierwalt2017}. We set out to use the morphology and kinematics of the HI to determine if the group is gravitationally bound. In the process, we discovered two new dwarf satellites via their HI content. 

The organization of this paper is as follows: Section \ref{groupdescription} describes the known properties of dm1623, an isolated, compact group of only dwarf galaxies. Section \ref{data} describes the reduction of the VLA and Gemini data that led to the discovery of two additional satellite group members. Section \ref{results} presents the resulting HI maps and optical spectra. In Section \ref{discbound}, we discuss evidence for whether the dm1623 dwarfs represent a gravitationally bound group, including a comparison to the properties of compact dwarf groups found in hydrodynamic simulations. In Section \ref{smfsection}, we compare the mass distribution of group members in dm1623$+$15 to the satellites around other dwarf hsots, and in Section \ref{disclss} we discuss the placement of this dwarf group within the surrounding large scale structure. Finally, in Section \ref{conc}, we summarize our conclusions. 

\section{TNT Dwarf Group dm1623+15}\label{groupdescription}
\begin{figure}[h]
  \centering
{\includegraphics[width=0.4\textwidth]{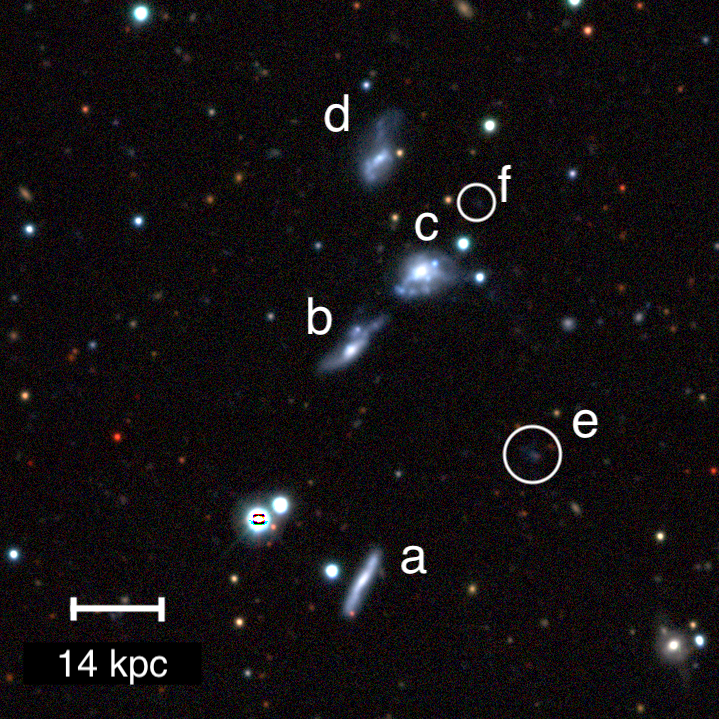}}
  \caption{Ground-based view of TNT Dwarf Galaxy Group dm1623+15: broadband optical grz image from the Legacy Sky Survey. Newly identified group members discovered via their HI content, 1623e and 1623f, are circled.\label{1623optical}}
\end{figure}

The two southernmost galaxies in the dwarf group dm1623+15, labeled a and b in Figure \ref{1623optical}, were originally identified in the TNT isolated, interacting dwarf pair sample \citep{Stierwalt2015}. The TNT Survey identified pairs of low mass ($10^{7} M_{\odot} < M_* < 5 \times 10^{9} M_{\odot}$) galaxies from SDSS with redshifts $0.005 < z < 0.07$, projected pair separations of $<$ 50 kpc, and velocity separations of $<$ 300 km/s. An additional isolation criterion required that all TNT pairs be more than 1.5 Mpc away from a massive host galaxy ($M_* > 5 \times 10^{9}$ M$_{\odot}$). The original pair, dm1623a and dm1623b, have a projected radial separation of 41 kpc, a projected velocity separation of 235 km/s, and a stellar mass ratio of $\sim$1.5:1. 

The higher mass galaxy in the original pair, dm1623a, has an estimated stellar mass of log(\Mstellar/\Msun) = 9.29 (see Section \ref{mstarcalc}) and, from \cite{Brinchmann2004}, a metallicity of Z = 0.008 and a star formation rate of SFR = 0.50 \Msun/yr. The lower mass galaxy in the original pair, dm1623b, has  log(\Mstellar/\Msun) = 9.10,  Z = 0.004, and SFR = 0.70 \Msun/yr. 

An additional two galaxies, dm1623c and dm1623d, were identified as potential group members due to their projected separation. Spectroscopic followup with the Dual Imaging Spectrograph (DIS) at the Apache Point Observatory revealed that dm1623c and dm1623d were at a similar redshift to the original pair \citep{Stierwalt2017}. 
Although not one of the originally identified pair members, dm1623c is the group member with the highest stellar mass (see Table 1). 

The final two group members, dm1623e and dm1623f, are barely visible in the Legacy Sky Survey images (\cite{deylegacy}; see Figure \ref{1623optical}) but were detected via their gas content as described in the following sections. 

\section{Observation and Data Reduction}\label{data}
\subsection{VLA Data}
The dwarf group dm1623+15 was observed with the Very Large Array from August 2022 to April 2024 in B/C/D-array configurations with integration times of 14.6/4.4/1.2 hours respectively. The observations in all three configurations had a total bandwidth of 8 MHz over 256 channels with a central frequency of 1373.8844 MHz. These correlator choices resulted in a channel width of 31.25 kHz or 4.5 km/s and velocity coverage from 9,562 km/s to 10,724 km/s.

Calibration and data reduction were carried out with the CASA software package \citep{CASA} using standard procedures. Manual flagging was performed to remove bad time ranges in the data and channels lost to radio frequency interference (RFI). Ultimately $\sim$15\% of the data were removed by flagging. The quasar 3C286 was observed at the start and end of each observing session to calibrate the bandpass and to set the flux density scale. The phase calibrator J1609+2641 was visited every 10-15 minutes throughout each observing session and used for complex gain calibration.

Imaging was performed with the CASA task \texttt{tclean}. Natural weighting was chosen to improve our sensitivity to low surface brightness emission. The radio continuum was subtracted using emission-free channels on the low and high ends of the frequency range. The final combined map with data from all three configurations has a synthesized beam of $\sim$7.5\arcsec~$\times$ 6.5\arcsec ($\sim$5 kpc at the group distance of 145 Mpc), with a noise level of $\sim$0.18 mJy beam$^{-1}$.

\subsection{Gemini Data}
A single long-slit spectrum, composed of six observations of 900 seconds each, was taken with GMOS on Gemini North in September 2024. The GMOS long slit spectrum was observed with the R400-G5305 grating for a resolution of R$\sim$1900 and a resulting wavelength range of $\sim$500-900 nm.  
The goal of the observations was to confirm group membership and associate the HI knots with the optical light in the Legacy Survey images. 
The 1.5\arcsec optical slit was oriented at a position angle of 192 degrees to include both 1623e and 1623f but not the other four galaxies.
The data were reduced using the DRAGONS software \citep{dragons} including bias subtraction, flat fielding, interpolation across chip gaps, removal of cosmic rays and bad pixels, wavelength calibration using arc lamp exposures, and flux calibration using the spectrophotometric standard star G191B2B. 




\subsection{Stellar Masses \& Luminosities}\label{mstarcalc}
We derive stellar masses for each of our dwarf galaxy group members using photometry from the Legacy Survey and the assumed group distance of 145 Mpc. Cutout images in the $g$ and $r$ bands were downloaded from the archive. We created custom apertures that included all emission above the level of 2 $\times$ RMS in the $r$-band image. Special care was taken to remove background galaxies and foreground stars when appropriate. Galactic extinction was estimated and removed using the IRSA online tool\footnote{https://irsa.ipac.caltech.edu/applications/DUST/} which uses the \citep{Schlafly2011} dust maps. The resulting $g$-band luminosities and $g-r$ colors were then used in combination with prescriptions for mass-to-light ratios derived specifically for dwarf galaxies \citep{masstolight} to estimate stellar mass. Mass-to-light ratios estimated from these SED-derived prescriptions are expected to be accurate to within 0.3 dex \citep{conroyML}.

We further estimate the B-band luminosities (L$_B$) for each dwarf and for the group using the same $g$ and $r$ band photometry. We use the following conversion from \cite{bell03}:\\
\indent$m_B = m_g + 0.2354 + 0.3915(m_g - m_r - 0.6102)$\\
Our stellar masses, B-band luminosities, and their associated uncertainties are reported in Table 1.

\section{Results}\label{results}
\subsection{Dwarf-Dwarf Interactions \& Two New Group Members Revealed by HI Maps}\label{HImasscalc}
\begin{figure}[!tbp]
  \centering
{\includegraphics[width=0.4\textwidth]{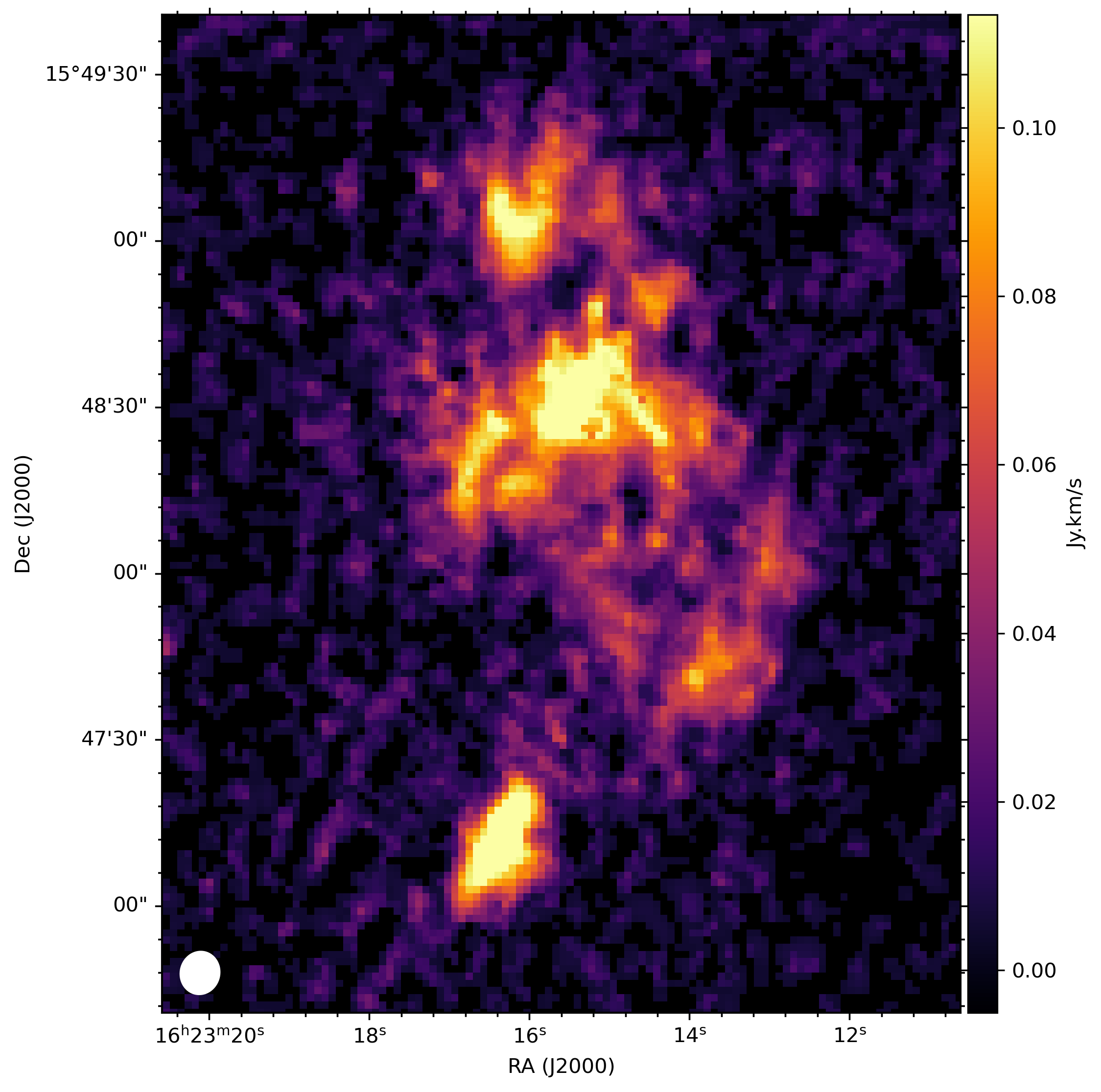}}
  \hfill
{\includegraphics[width=0.4\textwidth]{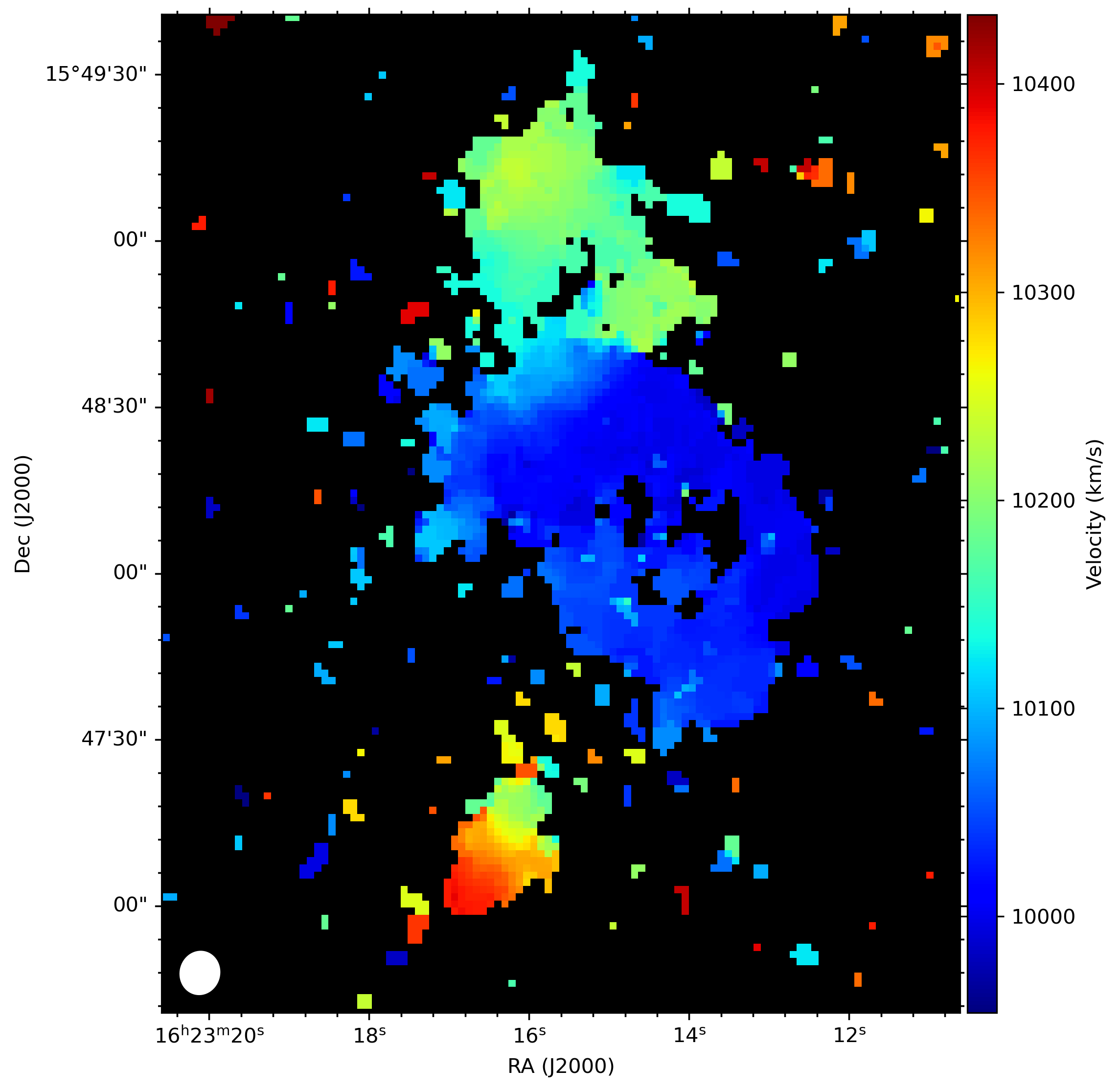}}
  \caption{HI intensity (moment 0) and HI velocity field (moment 1) for TNT dwarf galaxy group dm1623. The synthesized beam of $\sim$7.5\arcsec~$\times$ 6.5\arcsec~is shown in the lower left-hand corner.\label{momentmaps}}
\end{figure}

The final combined moment 0 (integrated line flux) and moment 1 (velocity) maps with data combined from all three array configurations are shown in Figure \ref{momentmaps}. Both moment maps were created using the \texttt{immoments} task in CASA and have a synthesized beam of $\sim$7.5\arcsec~$\times$ 6.5\arcsec ($\sim$5 kpc at the group distance of 145 Mpc). The HI intensity map was made using a flux density cutoff of 0.4 mJy beam$^{-1}$ (2$\sigma$), and the HI velocity map was made with a flux density cutoff of 0.6 mJy beam$^{-1}$ (3$\sigma$).

\begin{figure}[h!]
\begin{center}
\includegraphics[width=0.45\textwidth]{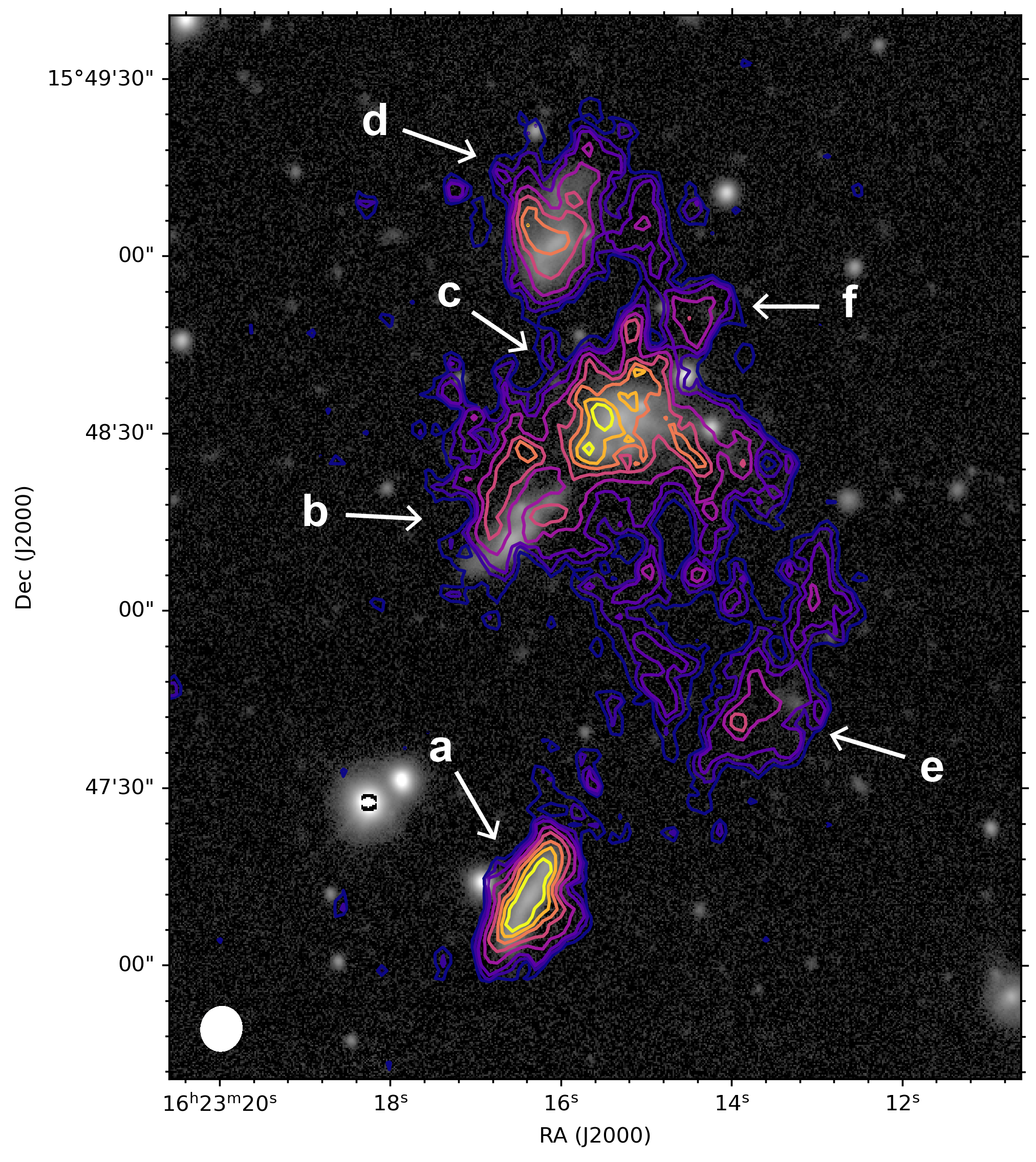}
\end{center}
\caption{Discovered Low Mass Satellites. The HI morphology of the isolated dwarf group dm1623 overlaid on a g-band optical image of the system. The HI contours are at the levels of 7.8, 10.4, 13.0, 18.2, 23.4, 28.7, 33.9, 39.1 $\times$ 10$^{20}$ atoms cm $^{-2}$, and the size of the restoring beam is indicated with the filled ellipse in the lower right hand corner. The arrows indicate the six dwarf group members with the lettering scheme adopted in this work. The newly discovered low mass satellites (e and f) are easily identified via their gas content but are only barely visible in the optical image.}\label{groupfig}
\end{figure}

The four brightest spots of emission in the HI intensity map match the positions of the previously known optical galaxies dm1623a-d. The map also reveals a significant amount of emission to the southwest of the group, as well as a knot of emission to the northwest of dm1623c. Both of these HI detections are coincident with previously unknown optical galaxies that are just above the detection limits of the Legacy Survey images  with m$_g \sim 22.32$ for 1623e and m$_g \sim 24.17$ for 1623f \citep{deylegacy}. The HI intensity map is overlaid as contours on top of the optical images in Figure \ref{groupfig}.

HI parameters including integrated flux density, central velocity, and velocity width, are presented in Table 1. To calculate HI masses, we made individual HI intensity maps for each galaxy, including only the channels with detected emission above a cutoff of 0.36 mJy beam$^{-1}$ which is 2$\times$RMS. The emission within the 2-$\sigma$ contour was then summed and divided by the pixel area to correct for the beam size and 
to determine the integrated flux density. The HI mass was then determined from the equation: $M_{HI}=2.36\times 10^5D^2S_{int}$, where D is the distance in Mpc and S$_{int}$ is the total HI line flux in Jy km/s. The HI masses for 1623b and 1623c are the most uncertain as they are heavily intertwined with each other. An aperture set by the 2$\sigma$ contour around the whole group includes log(M$_{HI}$/M$_{*}$) $=$ 10.40. This total HI mass suggests that M$_{HI}\sim  $9$\times$10$^{9}$~M$_{\odot}$ (36\% of the total) sits outside the galaxies.  The total HI mass recovered by the VLA is comparable to the HI mass of log (M$_{HI}$/M$_{*}$) $=$ 10.29 measured via single dish observations with the Arecibo Telescope \citep{Stierwalt2017}. 

In addition to the newly discovered galaxies dm1623e and dm1623f, there are two HI knots without optical counterparts that rise at least 2$\sigma$ above the noise level in the Legacy Sky Survey images. The HI knots can be clearly seen in Figure \ref{groupfig} to the west of dm1623d (log(M$_{HI,knot}$/M$_{\odot}$) $=$ 8.5 $\pm$0.2) and to the northwest of dm1623e (log(M$_{HI,knot}$/M$_{\odot}$) $=$ 8.4 $\pm$0.2). The knots have HI masses roughly one-half that of dm1623f and twice the lowest detectable HI mass. 

\begin{figure*}
\includegraphics[width=.95\linewidth]{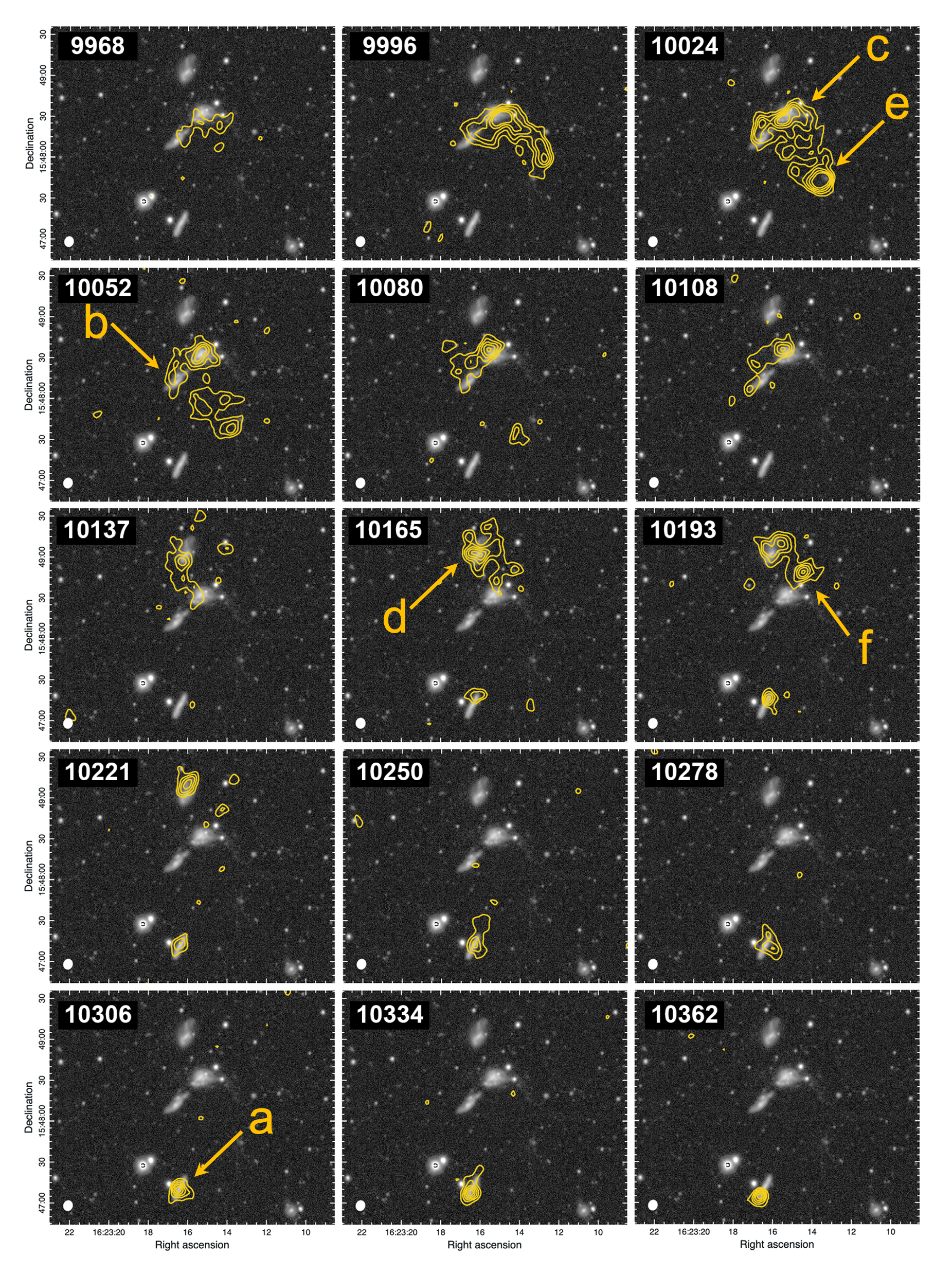}
\centering
\caption{Channel maps from the image cube resulting from the combined B, C, and D array data for dm1623. HI line emission is overlaid as contours on the g-band optical image. The contour levels are at 0.36 (2$\sigma$), 0.6, 0.8, 1.0, and 1.2 mJy beam$^{-1}$. The synthesized beam of $\sim$7.5\arcsec~$\times$ 6.5\arcsec~is shown in the bottom-left corner. Each dwarf group member is labeled in the channel map closest to its peak velocity (v$_{50}$).} 
\label{1623channelmaps}
\end{figure*}

The HI channel maps overlaid on the optical image of the group in Figure \ref{1623channelmaps} cover the full range of velocities for which HI emission is observed ($\sim$9954 km/s to 10376 km/s) in steps of 28 km/s.  The newly discovered group members, dm1623e and dm1623f, are most clearly seen at the channels for 10024 km/s and 10193 km/s respectively. Both new dwarfs have centralized HI peaks that make them clearly distinct systems while also being connected to the gas distributions of the original group members. As seen in the channel maps and the moment 1 map, the gas surrounding dm1623b, dm1623c, and dm1623e is most clearly connected in both 2D projected space and velocity space. To the north dm1623d and dm1623f are clearly kinematically connected to each other via their gas distributions and possibly connected to the rest of the group through gas near the center of the velocity range ($\sim$ 10,137 - 10,165 km/s). The southernmost galaxy dm1623a is the most kinematically and spatially distinct although it does contain gas at similar velocities to that found in dm1623d and dm1623f. 

\begin{table*}[h]
\centering
\footnotesize
{\bf{Table 1: Optical and HI Properties of the Dwarf Group Members}}\\
\begin{tabular}{l c c c c c c c r r c}
 \hline
 \hline
Name&RA&Dec&log(M$_*$)&log(L$_B$)&v$_{opt}$&S$_{int}$&v$_{50}$&W$_{50}~~~~$&log(M$_{HI}$)~~~~&log(f$_{gas})$\\
 &h:m:s&d:m:s&(M/\Msun)&(L/L$_{\odot}$)&(km s$^{-1}$)&(Jy km s$^{-1}$)&(km s$^{-1}$)&(km s$^{-1}$)~&(M/\Msun)~~~~&\\
 \hline

 1623a&16:23:16.37	&	15:47:11.48&9.27 ($\pm$0.15)&9.47& 10280 ($\pm$2.0) & 0.52 ($\pm$0.04)&10296 ($\pm$6)&198 ($\pm$15)&9.48 ($\pm$0.05)&0.21\\
 
 1623b&	16:23:16.57&15:48:12.00&9.09 ($\pm$0.10) &9.45&10031 ($\pm$2.0) &0.36 ($\pm$0.02)&10052 ($\pm$4)&140 ($\pm$10)&9.42 ($\pm$0.04)&0.33\\
 
 1623c&	16:23:15.31	&15:48:32.31&9.43 ($\pm$0.10)&9.80&10004 ($\pm$1.0) &1.15 ($\pm$0.03)&10020 ($\pm$5)&122 ($\pm$13)&9.73 ($\pm$0.04)&0.30\\
 
 1623d&16:23:16.07&15:49:02.43&9.06 ($\pm$0.10)&9.46&10139 ($\pm$1.0) &0.60 ($\pm$0.03)&10186 ($\pm$3)&106 ($\pm$7)&9.50 ($\pm$0.05)&0.44\\
 
 1623e&	16:23:13.26&15:47:44.42&7.07 ($\pm$0.10)&7.51& 10095 ($\pm$0.1)& 0.29 ($\pm$0.01)&10033 ($\pm$2)&40 ($\pm$4)&9.18 ($\pm$0.07)&2.11\\
 
 1623f&	16:23:14.26&15:48:50.37&6.10 ($\pm$0.20)& 6.80&10278 ($\pm$0.1)& 0.17 ($\pm$0.06)&10198 ($\pm$6)&98 ($\pm$14)&8.71 ($\pm$0.09)&2.61\\

Group& & & 9.84 ($\pm$0.30) & 10.18 & & & & & 10.40 ($\pm$0.05) & 0.56\\
 \hline
 \hline
\end{tabular}
\caption*{{Column (1): dwarf group member name; Columns (2) and (3): galaxy positions; Columns (4) and (5): Stellar mass and B-band luminosity calculated as described in Section \ref{mstarcalc}; Column (6): integrated flux density measured from HI intensity maps; Columns (7) and (8): central velocities and full width half max measured with Gaussian fits of the image cubes; Column (9): HI mass calculated as described in Section \ref{HImasscalc}; Column (10): gas fraction $f_{gas}=M_{HI}/M_*$.}\label{paramtable}}
\end{table*}

\subsection{Optical Velocities Confirm Group Membership of Two Newly Discovered Dwarfs}\label{gemsect}
\begin{figure}[!tbp]
  \centering
  {\includegraphics[width=0.5\textwidth]{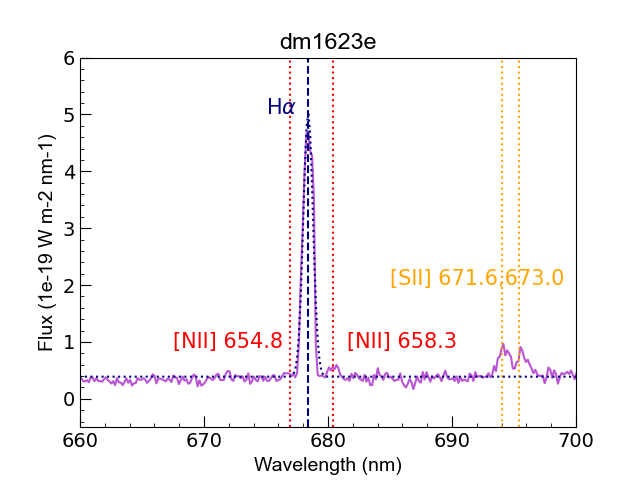}}
  \hfill
  {\includegraphics[width=0.5\textwidth]{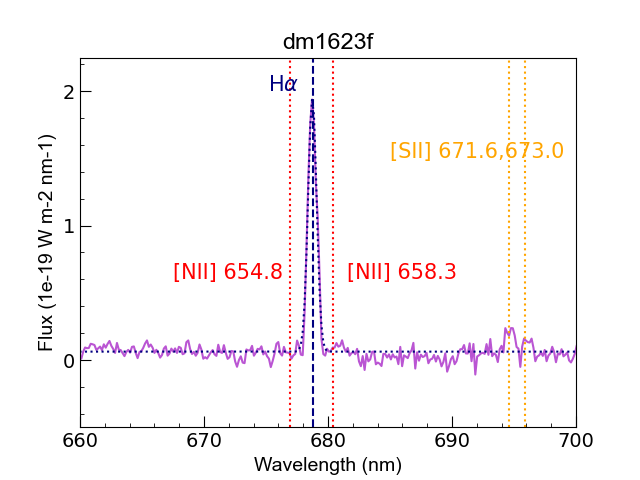}}
  \caption{Gemini spectra zoomed into the H$\alpha$ emission line (dashed blue line). The Gaussian fit used to determine the central wavelength is overlaid (dotted blue line). Two neighboring [NII] lines are noted by vertical red lines to show we do not expect blending to affect our redshift determinations. The sulfur doublet can be seen around 695 nm. \label{optspec}}
\end{figure}

To confirm the link between the HI knots detected in our maps and the very faint potential optical counterparts in the Legacy Survey imaging, we obtained GMOS optical spectra for dm1623e and dm1623f. The spectra, taken over a total integration time of 90 minutes, are shown in Figure \ref{optspec} zoomed in on the H$\alpha$ emission line.  The H$\alpha$ emission line is detected in both dwarf galaxies, placing them at the same distance as the previously known four group members. The resolution of the GMOS observations are sufficiently high ($\sim$0.3 nm) that we do not expect blending with the [NII] lines to significantly affect our redshift determinations. The optical velocities derived from the GMOS spectra are given in Table 1. 

The optical velocities differ from the central HI velocities by 62 km/s and 80 km/s for dm1623e and 1623f respectively. Although the central velocities are not a perfect match, the optical velocities are still within the range of velocities observed for the gas in each galaxy. The largest velocity differences in the group are for dm1623d, dm1623e, and dm1623f, which are the three galaxies with the lowest stellar masses. The galaxy with the smallest difference between its HI and optical velocity is dm1623a, the dwarf that appears to be least involved in the group member interactions. Thus these velocity differences could be a result of active interactions stirring up and removing the gas to a greater extent in the lower mass galaxies.

\subsection{Gas Fractions}
\begin{figure}[h!]
\begin{center}
\includegraphics[width=0.45\textwidth]{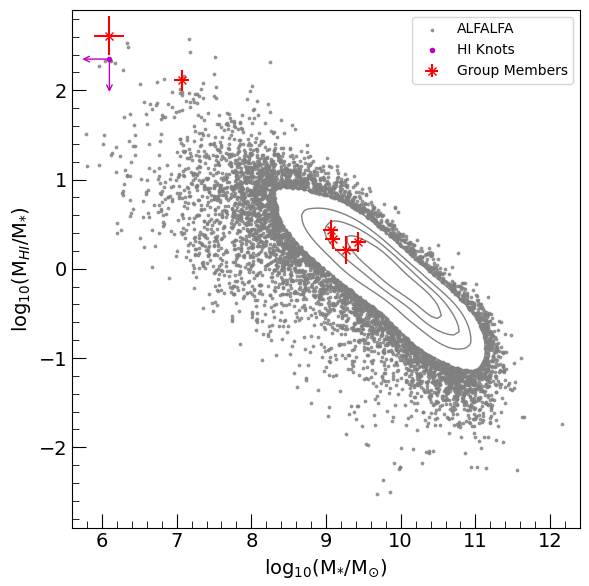}
\end{center}
\caption{Gas Fractions for dm1623$+$15 group members (purple stars) and for the group as a whole (blue star). The gray contours/points are HI-selected galaxies from the ALFALFA survey \citep{huangalfalfa}. Although much lower mass, the newly discovered satellites dm1623e and dm1623f are consistent in their gas-richness to the four more massive group members.}\label{gasfracfig}
\end{figure}

The two newly discovered satellites, dm1623e and dm1623f are 2-3 orders of magnitude lower in stellar and HI mass and have higher gas fractions than the original four group members. However, as shown in Figure \ref{gasfracfig}, their gas-richness (marked by red symbols) is consistent with the anti-correlation between stellar mass and $f_{gas} = M_{HI}/M_{*}$ demonstrated by the HI-selected galaxies from the ALFALFA survey \citep[gray contours and points;][]{huangalfalfa}. The ALFALFA galaxies are found across a range of environments so we might expect the isolated dm1623 dwarfs to be on the high end of the gas fraction range observed by ALFALFA. These two dwarf companions are also an order of magnitude smaller in both stellar and HI mass than the two companions discovered via HI observations around the spiral galaxy NGC 895 \citep{mightee}.

Also noted in Figure \ref{gasfracfig} by a set of purple arrows are upper limits on the stellar mass and gas fraction for the HI knots discussed in Section \ref{HImasscalc}. The HI knots have similar HI masses of log(M$_{HI}$/M$_{\odot}$) $=$ 8.4 and 8.5 but do not have optical counterparts detected above at least twice the background level in Legacy Survey images. Setting an upper limit to the stellar mass of each HI knot of log(M$_{*}$/M$_{\odot}$) $=$ 6.1, the stellar mass of dm1623f which is only barely detected in the same optical images, results in gas fractions of $\sim$2.35. If the stellar components of these HI knots are just below the sensitivity limits of our optical imaging, the high gas fractions of these isolated group dwarfs suggest that high resolution HI maps can be used instead to probe the satellite mass function.

\section{Discussion: Is this dwarf group gravitationally bound?}\label{discbound}
For a group of dwarf galaxies to eventually merge to form a single, larger galaxy, the group must be gravitationally bound. Studies of groups in simulations, where the true 3D structure is known, reveal up to 40\% of observed low mass galaxy groups are likely to be projection effects \citep{BeslaTNT}. For example, large transverse velocities could lead to relative 3D motions that are too high for the group members to be gravitationally bound, even when their line-of-sight velocities are similar. The line-of-sight velocity separations ($\Delta v_{los}$) between group members in dm1623$+$15 range from 2 km/s to 276 km/s. A $\Delta v_{los}$ of 276 km/s is on the high end of the range found in pair studies of massive galaxies \citep{patton2000galaxy,patton2011galaxy} and pair studies of dwarf galaxies \citep{Stierwalt2015}. However, this $\Delta v_{los}$ is still within those measured for bound groups in mock catalogs based on the Millennium-II simulation ($\Delta v_{los}$ up to 200-300 km/s for groups with a host of $9 < log(M_*/M_{\odot}) < 9.5$; \cite{sales13}). 

Simulations also find that dwarf-dwarf interactions that do not ultimately end in mergers (i.e. fly-bys) far outnumber those that result in coalescence \citep{Martin2021}. With the additional information provided by HI gas maps, we have more insight into the dynamics of the group and thus whether or not it is truly a gravitationally bound system. 

\subsection{Evidence of Interaction via HI Gas}

As shown most clearly in Figure \ref{1623channelmaps}, almost all of the dm1623$+$15 galaxies have bridges of gas that connect them to other group members. The channel maps provide the clearest evidence that the two more central galaxies, 1623b and 1623c, are in the process of merging. They share much of the same gas and all of their surrounding gas is at similar velocities. This pair further has the smallest projected separation of any two group members at 19 kpc. There is a clear gas bridge between dm1623c and the newly identified dm1623e spanning $\sim$9630 to 9750 km/s and a tentative detection of connecting gas between dm1623c and dm1623d from $\sim$9780 to 9910 km/s. The galaxies dm1623d and dm1623f are connected via gas that spans $\sim$9780 to 9900 km/s. dm1623a has a central HI velocity of 100 km/s higher than the rest of the group members, but its optical velocity is the same as that measured for dm1623f (see Table 1). dm1623a exhibits minimal gas interactions, but its HI profile spans a wide velocity range, overlapping with 1623d and 1623f.

Such gaseous bridges are typically signs of ongoing gravitational interactions between galaxies \citep{toomre1972galactic, bh92, YunM81}. Interacting dwarf galaxies in particular are known to sit in large gas reservoirs that are eventually re-accreted when not near a massive host \citep{pearson16}. Some compact groups of galaxies (sizes $\lesssim$ 100 kpc like dm1623$+$15) are found to have significant amounts of HI \citep[e.g.,][]{HCGwalker, HCGian}. That gas is initially found within the galaxies themselves but then later disperses to the intragroup medium as the evolution of the group progresses \citep{HCGjones}. Thus, the extended and inter-connected HI morphology of dm1623$+$15 suggests a gravitationally bound group hosting ongoing interactions between its members.

\subsection{The Dynamical Mass of dm1623$+$15}\label{dynloyal}
We can further test the likelihood of dm1623$+$15 being a physical group by estimating the amount of mass required for the six galaxies to be gravitationally bound.

Equating the 3D velocity dispersion ($\sigma_{3D}$) with the escape velocity for the group, and adopting the largest 2D distance between any group members  as the group size, gives a lower limit on the total (baryonic plus dark matter) mass required for the group to be gravitationally bound. We use the optical velocities for all six group members to calculate $\sigma_{3D} =$ 188 $\pm$1.3 km/s via:
\begin{equation} \sigma_{3D} = \sqrt{3}\times\sqrt{\langle v^2\rangle - \langle v \rangle^2} \end{equation}  

Using a velocity dispersion of 188 km/s and a group size of 69 kpc (between 1623a and 1623d), we estimate $\gtrsim$3.22$\times$10$^{11}$M$_{\odot}$ must be present in some form (baryonic plus dark) for the galaxies to be a gravitationally bound group. The observed baryonic mass alone (M$_* + $M$_{HI}$) is 3.20$\times$10$^{10}$M$_{\odot}$ so to reach this lower limit would require a ratio of $M_{dyn}/M_{bary} \gtrsim$10. This lower limit is a good match to the group of 5 dwarf galaxies discovered by \cite{paudel} with $M_{dyn}/M_{bary} = $23. Given the total B-band luminosity for dm1623$+$15, L$_B = 1.5\times 10^{10}$L$_{\odot}$, our lower limit on the total mass results in a mass-to-light ratio of M{$_{dyn}$}/L$_B \gtrsim$21 required for the group to be gravitationally bound. This mass-to-light ratio is consistent with the range of values of $\sim$1-100 observed for galaxy groups \citep{MLgroupshrad,MLgroupsshan} and for individual dwarf galaxies \citep{mateo98,MLursa,swaters11,spekkens14,MLsweet}. The lower limit on the total mass we estimate from the velocity dispersion (M$_{dyn}$ $\gtrsim$ 3.22$\times$10$^{11}$M$_{\odot}$) is also consistent with the total (baryonic$+$dark) mass expected for the group based on abundance matching. Using the stellar-to-halo mass relation from \cite{moster13} and a group stellar mass of log(M$_*$/M$_{\odot}$) $=$ 9.84, the group should have a mean halo mass of roughly 2.55$\times$10$^{11}$M$_{\odot}$ with a 1-$\sigma$ spread between 2$\times$10$^{11}$M$_{\odot}$ $\leq$ M$_{tot}$ $\leq$ 3.36$\times$10$^{11}$M$_{\odot}$.


Our quoted uncertainty in $\sigma_{3D}$ ($\pm$1.3 km/s) reflects only the statistical uncertainty due to the well-determined, optically-derived redshifts and does not account for systematic uncertainties. A large contribution to our systematic uncertainty is likely group membership so we investigate how $\sigma_{3D}$ would change if certain galaxies were not included. We calculate the same velocity dispersion (188 km/s) whether 1623e and 1623f are included or not. If 1623a, the most kinematically distinct galaxy, is excluded, the 3D velocity dispersion decreases to 168 km/s. Using instead this lower velocity dispersion and the 2D projected distance between dm1623d and dm1623e of 62 kpc as the size of the group, the resulting lower limit on the mass for the group to be gravitationally bound without dm1623a drops to $\gtrsim$2.02$\times$10$^{11}$M$_{\odot}$. This reduces the required M$_{dyn}$/L$_B$ for the group only slightly to M$_{dyn}$/L$_B \gtrsim$13.5.

Projected mass estimates are an alternative to the virial theorem for estimating the amount of mass in a galaxy group when only projected distances and line of sight velocities are known. We use the projected mass estimator method from \cite{projmass}:
\begin{equation}
M = \frac{f_{pm}}{G(N - \alpha)}\sum_i^N R_{p,i}\Delta V^2_i
\end{equation}
to estimate the total (baryonic + dark matter) mass of our group. Here $R_p$ is the 2D projected distance between an individual group member and the (unweighted) center of the group, $\Delta$V is the difference in the line-of-sight velocity between an individual group member and the mean group velocity, and $N$ is the total number of group members. We use $f_{pm} = 20/\pi$ and $\alpha = 1.5$ as adopted in the literature \citep{projmass, tully87}. 

When the sum is performed over all six group members, the projected mass measured for dm1623$+$15 is 6.65$\times$10$^{11}$M$_{\odot}$. If 1623a is excluded, the resulting projected mass is 5.23$\times$10$^{11}$M$_{\odot}$. Both of these estimates for the total group mass exceed  3.22$\times$10$^{11}$M$_{\odot}$, the lower limit on the total mass required for the group to be gravitationally bound, according to our estimate from the 3D velocity dispersion. These two estimates for the total (baryonic $+$ dark) mass for the group suggest mass-to-light ratios of M$_{dyn}$/L$_B \sim$44 and 35 which are again consistent with values observed for other groups \citep[e.g.,][]{MLgroupshrad,MLgroupsshan}. 

Finally, we can estimate the amount of dark matter in individual galaxies by fitting rotation curves to their gas velocity maps. Only one of the six galaxies, dm1623a, allows for a reasonable approximation of its rotation curve. The other galaxies are either too small to span at least three beam widths (dm1623b, dm1623e, and dm1623f) or are actively involved in an interaction that appears to be redistributing their gas (dm1623c and dm1623d). For dm1623a, we use the \texttt{FitTiltedRings} python code developed by \cite{Cooke2022} to fit a rotation curve to our 2D velocity map. The code performs a first order harmonic decomposition in concentric 2D tilted rings. The rings are spaced in steps of half of the FWHM or to include at least 30 pixels. The user then iterates on input values for the galaxy's dynamical center, inclination and position angles, and central velocity until arriving at the best fitting set of parameters. 

\begin{figure}
\centering
\includegraphics[width=.98\linewidth]{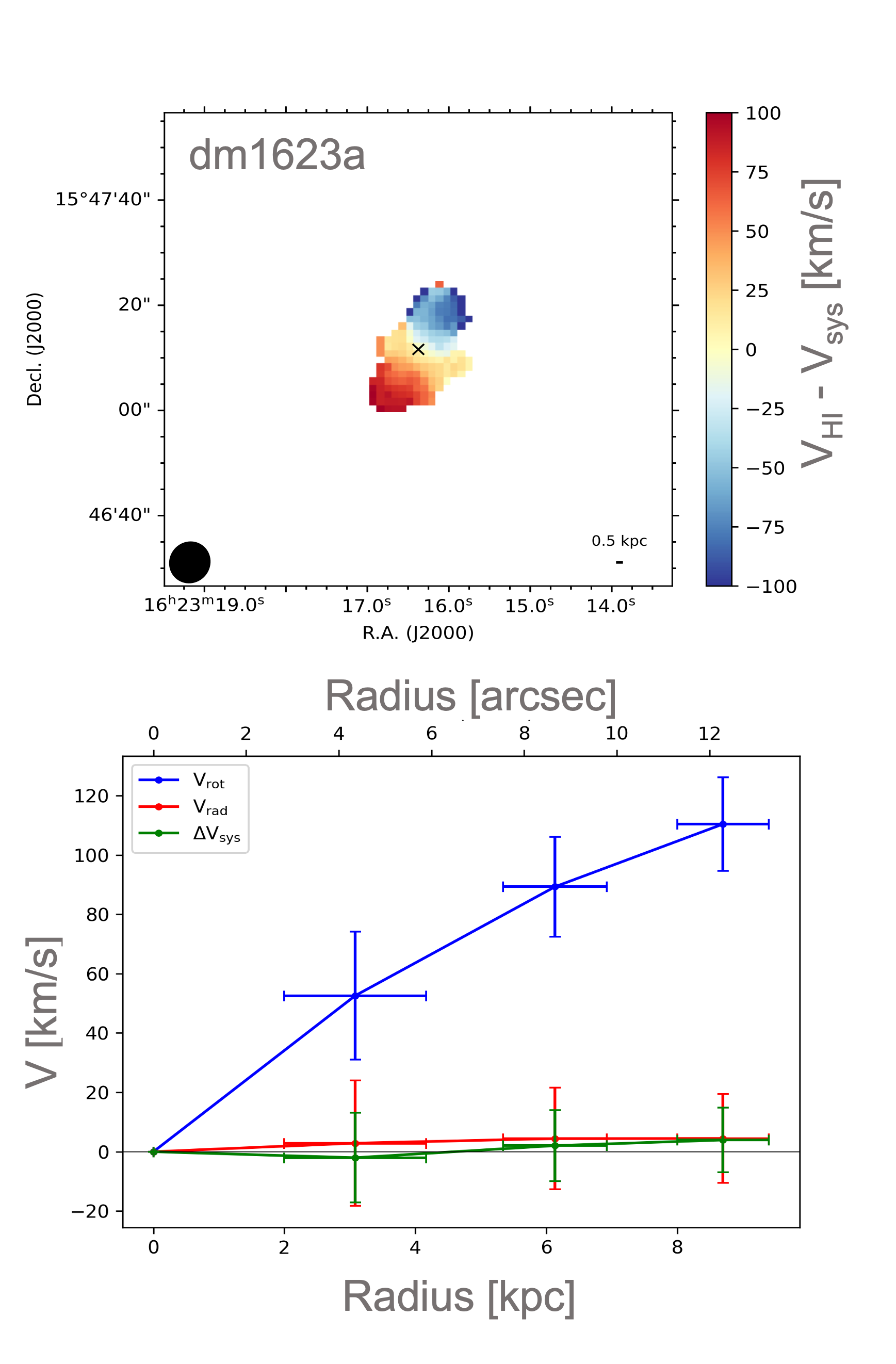}
\caption{Rotation curve fitting results for dm1623a. Top (input): 2D velocity map derived from the HI data cube for dm1623a. The color scale indicates how far the gas deviates from the central velocity of 10,285 km/s. The beam size is shown in the lower left corner. Bottom (output): Rotational velocity of the gas as a function of radius (blue curve). The fitting aims to minimize the fitted radial (line-of-sight) velocity (red curve) and the deviation from the systemic velocity (green curve). The rotation curve is still rising and thus the dynamical mass derived from it should be considered a lower limit.}
\label{rotcurves}
\end{figure}

The input 2D velocity map and the results from the rotation curve fitting code are presented in Figure \ref{rotcurves}. The dynamical mass derived from this rotation curve is log(M$_{dyn}$/M$_{\odot}$) $=$ 10.39 $\pm$0.3. Using the baryonic mass 
given in Table 1, this dynamical mass translates to 
a dynamical-to-baryonic mass ratio of M$_{dyn}$/M$_{bar}$ $=$ 5.0 $\pm$ 3.4. If we assume this ratio also applies to the group as a whole, we estimate a total (baryonic+dark) mass of 1.6 $\times$ 10$^{11}$M$_{\odot}$ which is a bit lower than the mass needed for the group to be gravitationally bound. However, the rotation curve does not extend far enough for the rotational velocity to level off and is clearly still rising. Thus, this estimate for the dynamical mass of dm1623a should be considered a lower limit.




\subsection{Comparison to Dwarf Groups in TNG50}

Further support for the possibility that dm1623$+$23 is a bound group is that its properties are a good match to those of the compact dwarf groups produced in the TNG50 simulation. TNG50 is a cosmological, magnetohydrodynamical simulation that produces $\sim$20,000 galaxies with M$_* >$ 10$^7$ M$_{\odot}$ \citep{tng50nelson,tng50pill}.  \cite{dgTNG50} specifically looked for compact dwarf groups (CDGs) produced by TNG50-1 and find they form with sizes $<$100 kpc and B-band mass-to-light ratios 10 $\leq$ M/L $\leq$ 100. These values are consistent with the size of dm1623$+$15 (69 kpc) and the mass-to-light ratios of the group (see Section \ref{dynloyal}). 

The 3D velocity dispersion determined for the group, 188 km/s, is higher than the highest velocity dispersion found for the TNG50 groups of 117 km/s. However, the velocity dispersions for the TNG50 groups are estimated differently than our 3D velocity dispersion estimate, specifically via: $\sigma = [(1/(N-1))\Sigma \Delta v_i^2]^{1/2}$ where $\Delta v_i$ is the difference between each galaxy's line of sight velocity and the group mean. Using this approach, we would estimate a velocity dispersion of $\sigma = $119 km/s for dm1623$+$15 which is still at the high end of the range for the TNG50 groups but consistent.

Also consistent between dm1623$+$15 and the TNG50 groups are the total group stellar mass, the projected mass estimate, and $\Delta$log(M$_*$)$_{max}$, the largest difference in stellar mass between any two group members. \cite{dgTNG50} find a range of $\Delta$log(M$_*$)$_{max}$ from $\sim$0.3 to 2.6. When we exclude dm1623f, which is below the stellar mass limit of the TNG50 work and thus would not be included in their sample, we find $\Delta$log(M$_*$)$_{max} = $2.36. 
Compared to the TNG50 CDGs, dm1623$+$15 is on the high end of the observed range of total group stellar mass at $[log(M_*/M_{\odot})]_{group} = $9.84. Significantly, many of the TNG50 CDGs host more than one dwarf with a mass greater than that of the SMC. We observe a gas fraction of $[M_{HI}/(M_{HI}+M_*)]_{group} =$ 0.78 which is high given the stellar mass of the group but well within the range observed for TNG50 CDGs (0.2 $< [M_{HI}/(M_{HI}+M_*)]_{group} <$ 0.9). 

According to \cite{dgTNG50}, most of the CDGs identified in the simulation at z $>$ 0.2 continue to undergo changes by either accreting new galaxies and/or undergoing mergers between the original group members. 
Most of the CDGs found at z $\lesssim$ 0.5 take between 1-3 Gyr to completely merge into a single galaxy. The authors also note that the groups are unlikely to break apart, suggesting that only a small fraction are transient, unbound structures.

The column densities and the velocity widths of the HI emission in the dwarf groups are also consistent with damped Lyman-$\alpha$ systems (DLAs) detected at higher redshift (e.g., \cite{DLA1, DLA2}.  Given dwarf groups could be more common at higher redshift, systems analogous to the one presented here may be responsible for some of the observed DLAs.

In summary, dm1623$+$15 appears to be a gravitationally bound group of only dwarf galaxies that may eventually merge to form a single, isolated system. The group does not have a clear host as there are three galaxies (dm1623a, dm1623b, and dm1623d) of roughly similar stellar and HI mass. dm1623c has slightly higher stellar and HI mass and is positionally central to the group. However, its central velocity is lower than all other group members so it does not appear to be at the dynamical center. Simulations suggest timescales for merging from 1-3 Gyr and, given the lack of a clear host, the merging time for  dm1623$+$15 may be on the longer end of this range.

\section{The Mass Distribution of Satellites in dm1623$+$15}\label{smfsection}
Recent efforts to determine the satellite mass function around dwarf hosts typically start with an isolated dwarf galaxy that clearly serves as the group host \citep{carlinMADCASH, delvemcnanna, garling21, davis21, carlinMADCASH24, dolivaMADCASH, medoffMADCASH, LiELVES}. 
Although there is no clear host in the dm1623$+$15 group, the most massive galaxy , dm1623c (M$_*$ $=$2.7$\times$10$^9$M$_{\odot}$), is a very near match to the stellar mass of the LMC (M$_*$ $=$2.6$\times$10$^9$M$_{\odot}$; \cite{dooley17}) and is the most likely candidate. Relative to the other group members, dm1623c is the most centrally located in projected space and more massive by at least 1.5 times. It is not, however, centrally positioned in velocity space, sitting instead at the low end of the velocity range covered by both the optical spectra and the HI maps.

\begin{figure}[h]
  \centering
{\includegraphics[width=\columnwidth]{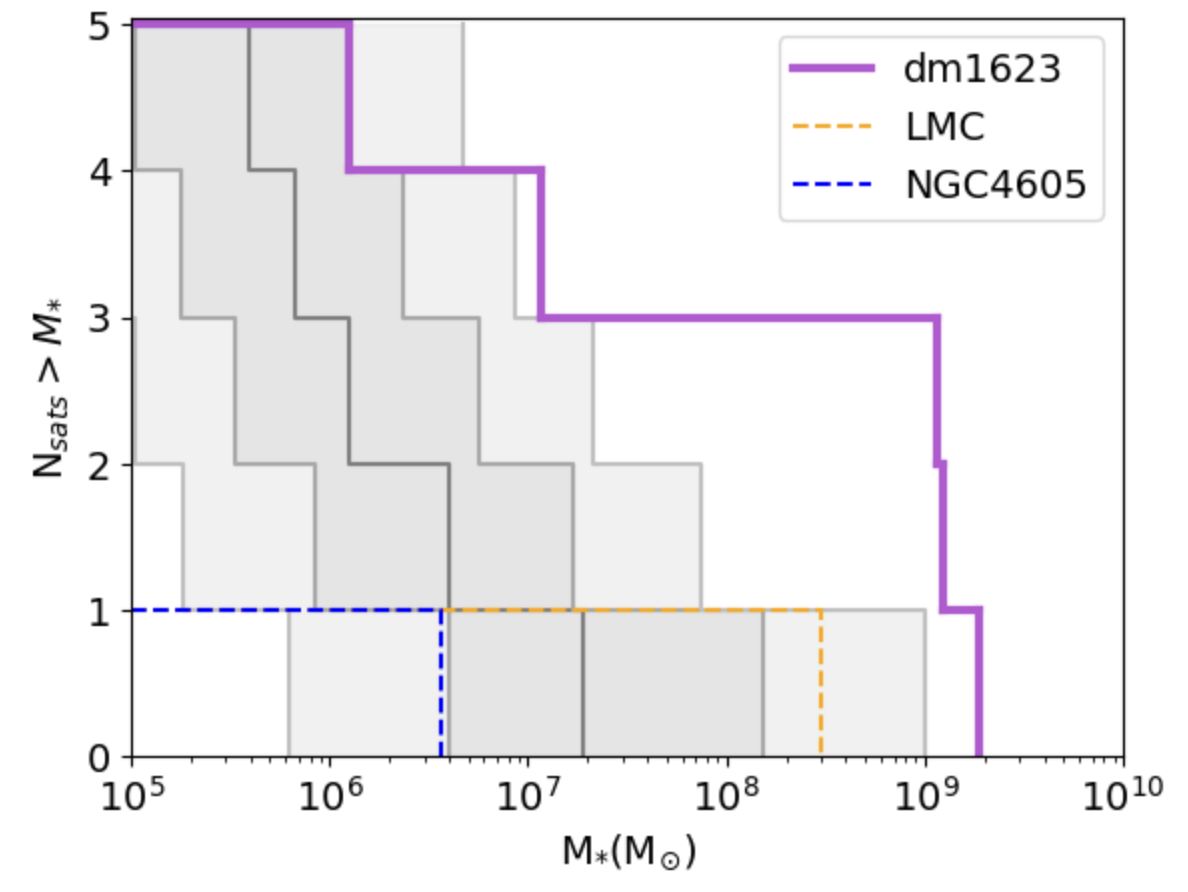}}
  \caption{The cumulative distribution function of the dwarf group members in dm1623$+$15 when setting dm1623c (log(M$_*$/M$_{\odot}$) $=$ 9.43) as the host (magenta line). Also shown are the cumulative distribution functions of satellites down to M$>$10$^5$M$_{\odot}$ around dwarf hosts of similar stellar mass: the single satellite around NGC 4605 \citep{LiELVES} and the SMC as a satellite of the LMC. The theoretical prediction of \citep{manwadkar} for a galaxy with log(M$_*$/M$_{\odot}$) $\sim$ 9.5 (gray line) is shown along with the 1$\sigma$ (68\%; light gray) and 2$\sigma$ (95\%; darker gray) confidence intervals. \label{smffig}}
\end{figure}

In Figure \ref{smffig}, we show the cumulative distribution function (CDF) of satellites around dm1623c (purple line). Also shown are the distributions of satellites around two other dwarf hosts with stellar masses similar to that of dm1623c. 
One satellite (blue dashed line) was discovered via deep optical imaging around NGC 4605, an isolated dwarf in the ELVES-DWARF survey (M$_*$$=$3$\times$10$^9$M$_{\odot}$; \cite{LiELVES}). Additionally, we show the SMC (M$_*$$=$3.1$\times$10$^8$M$_{\odot}$; \cite{snezSMC}) as the only potential satellite within 50 kpc of the LMC with a stellar mass $>$10$^5$M$_{\odot}$ (yellow dashed line). Finally, we show the predicted satellite mass function (dark gray line) for a host with M$_*$$=$3.2$\times$10$^9$M$_{\odot}$ determined from the stellar-to-halo mass relation (SHMR) of \cite{manwadkar} and as presented in \cite{LiELVES}. The SHMR uses the semi-analytical model GRUMPY \citep{grumpy} and the Caterpillar zoom-in simulations \citep{caterpillar}. The shaded regions indicate 1$\sigma$ (68\%; light gray) and 2$\sigma$ (95\%; darker gray) confidence intervals. We focus on these two groups for comparison since a strong link is observed between host mass and satellite abundance \citep{SAGA24Mao}. Comparisons across these satellite populations are complicated by varying completeness levels and by the fact that the LMC is not isolated. Thus, we do not make any corrections for completeness and use this plot for rough comparison only. 

Although a much larger group than NGC 4605 and the LMC in this mass range, the total number of satellites in dm1623$+$15 is within the range of the predictions: both the purple and dark gray lines reach a total of five satellites around 10$^6$M$_{\odot}$.
However, the distribution of the satellites in dm1623$+$15 is very ``top-heavy''. The masses of the dm1623$15$ satellites are overall much higher and three of the five satellites have masses M$_*$ $>$ 10$^9$M$_{\odot}$ or within a factor of $\sim$1/2 of the host mass. A similar top-heaviness was observed for a group of five dwarfs discovered via the similarities in their line-of-sight velocities \citep{paudel}. The most massive dwarf in that group (M$_*$$=$2.8$\times$10$^8$M$_{\odot}$) is surrounded by four companions with stellar masses 1.5$\times$10$^7$M$_{\odot}$ $<$ M$_*$ $<$ 1.9$\times$10$^8$M$_{\odot}$. Given the lack of a clear host, it is possible that the dm1623$+$15 group is an outlier in its satellite mass function because it is two or more hosts coming together to form a single system. 

We also cannot rule out the possibility that the two newly discovered dwarf galaxies are tidal dwarf galaxies given that they are embedded in the extensive tidal debris that surrounds the group. Using the velocity widths and baryonic masses given in Table 1, dm1623a, dm1623b, and dm1623e all lie within one standard deviation of the baryonic Tully Fisher relation as parameterized for gas-rich, dwarf galaxies \citep{tracTF,starkTF}. dm1623c and dm1623d both lie within two standard deviations of the relation, suggesting they too have M$_{dyn}$/M$_{bar}$ ratios comparable to other low mass galaxies. Only dm1623e lies significantly above the relation with a much lower observed rotation velocity (20 km/s) than predicted (74 km/s) given its baryonic mass. If this lower rotational velocity is due to a lack of dark matter, it would suggest an M$_{dyn}$/M$_{bar}$ roughly 13 times lower in dm1623e than for other gas-rich dwarfs.
Such low levels of dark matter could suggest this dwarf is galaxy formed from material tidally removed due to the interactions between the more massive dwarfs \citep{ducTDG, jacksonDM}. If the two newly discovered dwarf galaxies, dm1623e and dm1623f, are tidal in nature, that would even further skew the observed satellite mass function away from theoretical predictions. 

\begin{figure}[h]
  \centering
{\includegraphics[width=\columnwidth]{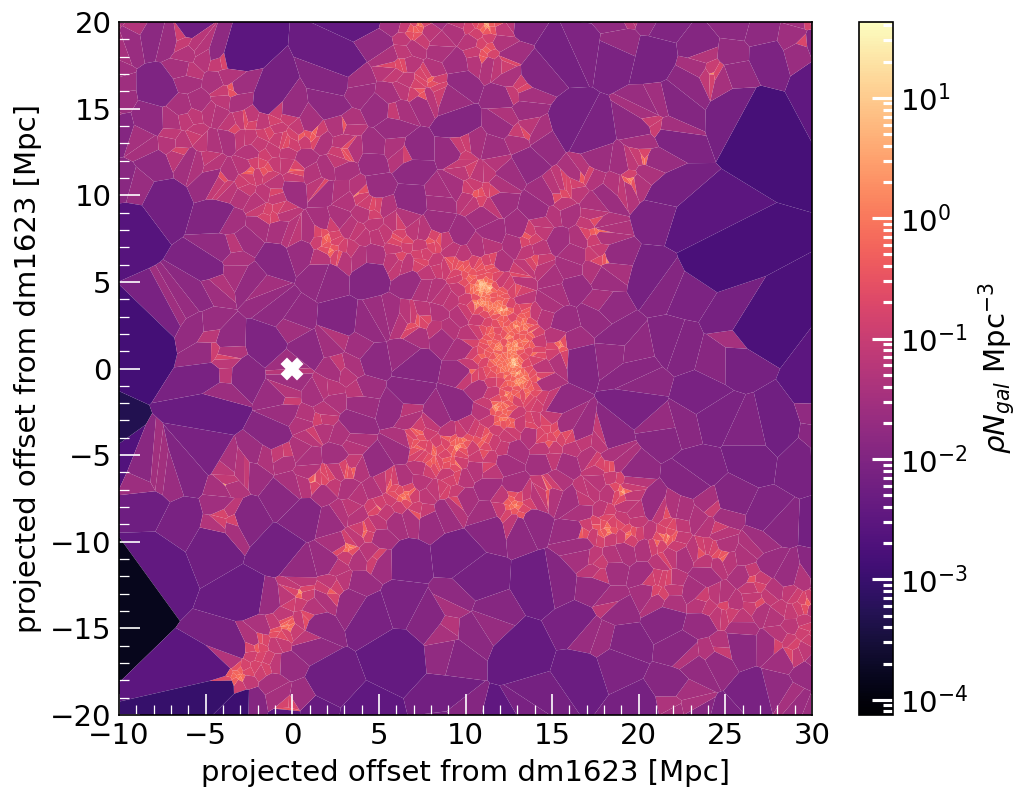}}
  \caption{The projected Voronoi tessellation of galaxies in a redshift slice (0.0315 $< z <$ 0.0365) around dm1623$+$15. Each Voronoi cell corresponds to a galaxy projected onto the plane of the sky and is colored by the local galaxy density, calculated as the inverse of the cell area scaled by the depth of the redshift slice. The white ``X" marks the position of dm1623$+$15, and the axes are projected distance to dm1623$+$15.\label{lss}}
\end{figure}

\section{The Large-scale Environment of dm1623$+$15}\label{disclss}

\quad Both observations and simulations show that the universe is threaded with a filamentary structure known as the ``cosmic web'' \citep[e.g.,][]{illustris,CWlibeskind,cwsdss}. The details of how galaxies form and evolve within this large scale network are still being worked out. Low redshift studies have found that some galaxy properties, like color and stellar mass, are well correlated with distance to the nearest cosmic web filament \citep[e.g.,][]{kuutma17,laigle18,luber19}. The link between a galaxy's position in the cosmic web and its star formation is less clear. Some studies find more passive and/or massive galaxies closer to filaments \citep{kraljic18} while others observe a so-called ``cosmic web enhancement'' with filament galaxies exhibiting increased star formation or delayed quenching \citep{Vulcani19,Kotecha22}. 

The impacts of the cosmic web, specifically impacts on the cold gas reservoirs of galaxies, are also likely to be felt differently at different stellar mass \citep[e.g.,][]{conselice13,sanchez14}. For example, \cite{odekon18}, using data from the ALFALFA Survey \citep{giovanelli05}, found that galaxies in the stellar mass range $10^{8.5} < M_{*} < 10^{10.5} M_{\odot}$ were increasingly HI deficient when closer to a filament spine. However, using stacked data from the HIPASS survey, \citet{kleiner17} found galaxies with $M_{*} \geq 10^{11} M_{\odot}$ had higher HI fractions when found within 0.7 Mpc of a filament spine than those more than 5 Mpc away. 

\quad To consider the effect of the cosmic web on dm1623$+$15, we reconstruct the surrounding large scale density field by performing a Voronoi tessellation using galaxies with spectroscopic redshifts available from SDSS in the immediate environment. Voronoi tessellations are a geometric tool used to partition space into regions around a set of discrete points, such that each region contains all locations closer to its associated point than to any other. In cosmology, Voronoi tessellations are widely used to quantify the local environment of galaxies by dividing the cosmic web into regions of influence centered on individual galaxies \citep[e.g.,][]{VTramella, VTvavilova}.

\quad In Figure \ref{lss}, we present the Voronoi tessellation for the region around dm1623$+$15, constructed using all SDSS galaxies within the redshift range 0.0315 $< z <$ 0.0365. This redshift range corresponds to a line-of-sight depth of approximately 20 Mpc. From Figure \ref{lss}, it is clear that dm1623$+$15 does not closely overlap with a filament such as the two that intersect at a bright overdensity at (10,0) Mpc. Instead dm1623$+$15 lies on a faint over density, or tendril, that stretches from the west side of that overdensity and through the neighboring underdense void.

\quad Cosmic web tendrils are small, filamentary structures embedded within voids \citep{VTalp}. \cite{odekon18} found galaxies in this tendril environment to be more HI-rich than their counterparts of similar stellar mass and local density that are found closer to filaments. They further found galaxies in tendrils to be more massive than true void galaxies. Together these  observations present a scenario in which tendrils are an ideal location for gas-rich mergers between dwarf galaxies within bound groups to contribute to the build up of more massive galaxies. The star-forming, HI-rich dwarf galaxies of the bound dm1623$+$15 group provide an excellent example of this evolutionary scenario in action. 

Compact dwarf groups in simulations are also found mostly in densities of just over 25th percentile, i.e. intermediate density environments \citep{dgTNG50}. Those authors suggest there may be a minimum density required for dwarf groups to form, but placed in too dense of an environment, the dwarfs are more likely to be accreted by a massive host.

\section{Conclusions}\label{conc}
We present our analysis of the gas content of the isolated, compact dwarf galaxy group dm1623$+$15 using HI maps taken with the VLA in B$+$C$+$D configurations. This dwarf group was identified in TiNy Titans \citep{Stierwalt2017} and is thus $>$1.5 Mpc from a massive host (M$_* >$5$\times$10$^9$M$_{\odot}$). We discover two new HI-rich group members not readily identified in the optical, determine the likelihood that the group is physically bound, compare to simulations of compact dwarf groups, and investigate the group's position in the cosmic web. We find:

\begin{enumerate}[(i)]
    \item The group of six dwarf galaxies has a total HI mass of 10$^{10.4}$M$_{\odot}$ that spans $\sim$400 km/s. Approximately 36\% of that total resides outside of the galaxies. 
    \item Two newly discovered dwarf galaxy satellites barely visible in Legacy Survey images are easily identified via their HI content. They have masses log(M$_*$/M$_{\odot}$)$=$7.07 with log(M$_{HI}$/M$_{\odot}$)$=$9.18 for 1623e and log(M$_*$/M$_{\odot}$)$=$6.10 with log(M$_{HI}$/M$_{\odot}$)$=$8.71 for 1623f. We confirm the association between the HI and the optical counterpart with Gemini optical spectra. 
    \item The HI maps also reveal two HI knots (log(M$_{HI}$/M$_{\odot}$)$=$8.5$\pm$0.2 and 8.4$\pm$0.2) without optical counterparts. If the stellar components are just below the sensitivity limits of our optical imaging, the high gas fractions suggest that HI maps can be used instead to probe the satellite mass function in isolated dwarf groups.
    \item The HI map shows clear signs of interaction between five of the six galaxies, including the two newly discovered dwarf satellites. The remaining galaxy, dm1623a, does not appear to be physically connected but contains gas at similar velocities to the other group members. 
    \item The projected mass estimate for the group is 6.65$\times$10$^{11}$M$_{\odot}$ which is greater than the mass required for the group to be gravitationally bound (3.22$\times$10$^{11}$M$_{\odot}$) as calculated from the 3D velocity dispersion of 188 km/s. The masses and kinematics of the group members, along with the clear signs of interaction in the HI map, indicate the group is a gravitationally bound entity.
    \item We estimate the dynamical mass for dm1623a using a rotation curve fit by a 2D tilted ring model. The resulting dynamical mass log(M$_{dyn}$/M$_{\odot}$)$=$10.39 translates to M$_{dyn}$/L$_B$ $=$ 8.3 $\pm$ 6.4 and M$_{dyn}$/M$_{bary}$ $=$ 5.0 $\pm$ 3.4. The rotation curve is still clearly rising which suggests these values should be considered lower limits. 
    \item The dm1623$+$15 group is comparable to the compact, dwarf groups found in the TNG50 hydrodynamic simulation. The group has a size (69 kpc), mass-to-light ratio (M$_L$/B$\sim$44), and gas fraction (f$_{gas} = $ 0.78) 
    that are consistent with the groups found to persist at the current epoch in the simulation. These groups typically take 1-3 Gyr to merge into a single galaxy.
    \item The satellite mass function for the dm1623$+$15 group is consistent with predictions from simulations in the total number of satellites but is more top-heavy. dm1623 may instead be two or more dwarf groups coming together. 
    \item A Voronoi tessellation constructed using SDSS galaxies reveals the group is located outside of an intersection between two filaments. This intermediate density tendril environment may be an ideal location for gas-rich mergers between dwarf galaxies to contribute to the build up of more massive galaxies. 
\end{enumerate}

\section*{Acknowledgements}\label{acknow}

S. Stierwalt and Z. Goldberg Little gratefully acknowledge support from the Occidental College Undergraduate Research Center and Norris Foundation Science Scholars Fund. They also thank E. Momjian for the extensive help during our visit to Socorro with the reduction of the VLA data. S. Stierwalt thanks C. Martinez-Vazquez and T. Mocnik for their support in planning and executing the Gemini observations, and K. Labrie for the support in analyzing the unusual dataset. 

This work has made use of the NASA/IPAC Extragalactic Database (NED) which is operated by the Jet Propulsion Laboratory, California Institute of Technology, under contract with the National Aeronautics and Space Administration. 

\software{CASA (CASA Team et al. 2022), DRAGONS (Labrie et al. 2023), CARTA (doi:105281/zenodo.3377984)}

\bibliography{HIreferences}{}
\bibliographystyle{aasjournalv7}
\end{document}